\documentclass[aps,preprint]{revtex4}%
\usepackage{amssymb}
\usepackage{amsfonts}
\usepackage{amsmath}
\usepackage{graphicx}%
\providecommand{\U}[1]{\protect\rule{.1in}{.1in}}
\newtheorem{theorem}{Theorem}
\newtheorem{acknowledgement}[theorem]{Acknowledgement}

\begin{document}
\title{Teleparallel equivalent of higher dimensional gravity theories}
\author{N. Astudillo-Neira}
\email{nastudillo@udec.cl}
\affiliation{Departamento de F\'{\i}sica, Universidad de Concepci\'{o}n, Casilla 160-C,
Concepci\'{o}n, Chile.}
\author{P. Salgado}
\email{pasalgad@udec.cl}
\affiliation{Departamento de F\'{\i}sica, Universidad de Concepci\'{o}n, Casilla 160-C,
Concepci\'{o}n, Chile.}
\date{March 17, 2017}

\begin{abstract}
The equivalence between the Lanczos-Lovelock and teleparallel gravities is
discused. It is shown that the teleparallel equivalent of the Lovelock gravity
action is generated by dimensional continuation of the teleparallel equivalent
of the Euler characteristics associated to all the lower even dimensions. It
is also found that the teleparallel equivalent of the $(i)$ $d$-dimensional
Euler characteristic is a closed form and gauge invariant, $(ii)$ Lovelock
action are invariant both under the Poincare group and diffeomorphisms.

\end{abstract}
\maketitle

\section{Introduction}

Teleparallel gravity can be interpreted as a gauge theory for the translation
group \cite{eins,unz,per,per1,andr}. Due to the peculiar character of
translations, any gauge theory including these transformations will differ
from the usual internal gauge models in many ways, the most significant being
the presence of a tetrad field. On the other hand, a tetrad field can be
naturally used to define a linear Weitzenb\"{o}k connection, which is a
connection presenting torsion, but no curvature. A tetrad field can also be
naturally used to define a riemannian metric, in terms of which a Levi-Civita
connection can be constructed. As is well known, it is a connection presenting
curvature, but no torsion. It is important to keep in mind that torsion and
curvature are properties of a connection \cite{per}, and that many different
connections can be defined on the same space \cite{per1,andr}. Therefore one
can say that the presence of a nontrivial tetrad field in a gauge theory
induces both a teleparallel and a riemannian structure in spacetime. The first
is related to the Weitzenb\"{o}k, and the second to the Levi-Civita connection.

Teleparallel gravity is a gauge theory for the translation group
\cite{eins,unz,per,per1,andr} whose gauge connection for this group is given
by $A=e^{a}P_{a}.$ \ The corresponding covariant derivative is then given by
$D\phi=\left(  dx^{a}+e^{a}\right)  \partial_{a}\phi=h^{\text{ }a}\partial
_{a}\phi$ and the 2-form torsion is given by $T^{a}=K_{\text{ \ }c}^{a}h^{c},$
where $h^{\text{ }a}=Dx^{a}=dx^{a}+e^{a}$ is the called tetrad field (see
Appendix A) and $K^{ab}$ is the so called one-form contorsion. \ The $(3+1)$-
dimensional action for teleparallel gravity, also known as the teleparallel
equivalent\textbf{, }given by Eq. ($14$) of Ref. \cite{per1} can be then
written in the form (see also \cite{sal1,sal2})
\begin{equation}
S_{Tel}=-\int\varepsilon_{abcd}K_{l}^{a}K^{lb}h^{c}h^{d}. \label{tel39}%
\end{equation}
Using the identity $D\varepsilon_{abcd}=0,$ it is possible to show that the
action (\ref{tel39}) can be written as
\begin{equation}
S_{Tel}=-\int\varepsilon_{abcd}K^{ab}T^{c}h^{d}. \label{tele39}%
\end{equation}
Since $\delta h^{\text{ }a}=0$ (see Appendix B), we have that the action
(\ref{tel39}) is invariant under local translations and by construction is
invariant under local Lorentz rotations and diffeomorphism. This means that
the action (\ref{tel39}) is genuinely invariant under the Poincar\'{e} group
and under diffeomorphism.

The purpose of this paper is to find the teleparallel equivalent of the
$d$-dimensional Euler classes and to show that the teleparallel equivalent of
the Lanczos-Lovelock gravity action is $(i)$ generated by dimensional
continuation of the Euler characteristics associated to all the lower even
dimensions; $(ii)$ is invariant under the Poincare group as well as diffeomorphisms.

This paper is organized as follows: In Section~2, we briefly review the
foundations of the so-called teleparallel gravity theory developed in Refs.
\cite{eins,unz,per,per1,andr}. In Section~3, we will make a short review about
the Lanczos-Lovelock gravity theory. Section 4 contains the results of the
main objective of this work, namely: to find the teleparallel equivalent of
the $d$-dimensional Euler characteristic and to show that the teleparallel
equivalent of the Lovelock gravity action is generated by dimensional
continuation of the Euler characteristics associated to all the lower even
dimensions. In this section it is also found that the teleparallel equivalent
of $(i)$ the $d$-dimensional Euler characteristic is a closed form and gauge
invariant; $(ii)$ some interesting particular cases related to the choice of
Lovelock lagrangian coefficients are invariant under the Poincare group as
well as under diffeomorphisms.\ We finish in Section~5 with some final remarks
and considerations on future possible developments.

\section{Teleparallel Gravity}

The fundamental field of a gauge theory is the connection. Objects with a well
defined behavior under local transformations are called covariant under those
transformations. It is a remarkable fact that usual derivatives of such
covariant objects are not themselves covariant. In order to define derivatives
with a well defined tensor behavior, it is necessary to introduce connections
$A_{\mu}$, which behave like vectors in what concerns the space-time index,
but whose non tensorial behavior in the algebraic indices just compensates the
non tensoriality of ordinary derivatives. \ 

Connections related to space-time groups, such as the Lorentz group, are
called linear connections, have a direct relation with spacetime because they
are defined on the bundle of linear frames, which is a constitutive part of
its manifold structure. These bundles have some properties not found in the
bundles related to internal gauge theories. \ For example its these bundles
exhibits soldering, which leads to the existence of torsion for every
connection \cite{koba}. Linear connections, in particular Lorentz connections,
always have torsion, while internal gauge potentials do not have. It should be
noted that a vanishing torsion is quite different from a non-existent or non
defined torsion.

A spin connection or Lorentz connection $A$ is a 1-form assuming values in the
Lie algebra of the Lorentz group, $A=\frac{1}{2}A^{ab}J_{ab}$, where $J_{ab}$
are the Lorentz generators. This connection defines the covariant derivatives
$D=d+A=d+\frac{1}{2}A^{ab}J_{ab}$, whose second part acts only on the
algebraic, or tangent space indices. \ In the case of soldered bundles, a
tetrad field relates Lorentz with external tensors. For example, if $\phi^{a}$
is an internal or Lorentz vector, then $\phi^{\rho}=h_{a}^{\text{ \ }\rho}%
\phi^{a}$ will be a spacetime vector.

On the other hand, a general linear connection $\Gamma_{\text{ \ }\nu\mu
}^{\rho}$ is related to the corresponding spin connection $A_{b\mu}^{a},$
through an identity, sometimes called Weyl's lemma:%
\begin{equation}
\tilde{\nabla}_{\mu}h_{\text{ \ }\nu}^{a}=\partial_{\mu}h_{\text{ \ }\nu}%
^{a}-\Gamma_{\text{ }\nu\mu}^{\rho}h_{\rho}^{\text{ \ }a}+A_{\text{ \ }b\mu
}^{a}h_{\text{ \ }\nu}^{b}=0,
\end{equation}
from where%
\begin{equation}
\Gamma_{\text{ }\mu\nu}^{\rho}=h_{a}^{\text{ \ }\rho}\left(  \partial_{\mu
}h_{\text{ \ }\nu}^{a}+A_{\text{ \ }b\mu}^{a}h_{\text{ \ }\nu}^{b}\right)
\equiv h_{a}^{\text{ \ }\rho}D_{\mu}h_{\text{ \ }\nu}^{a}. \label{veintinueve}%
\end{equation}
The inverse relation is, consequently,%
\begin{equation}
A_{\text{ \ }b\mu}^{a}=h_{\nu}^{\text{ \ }a}\partial_{\mu}h_{b}^{\text{ \ }%
\nu}+h_{\nu}^{\text{ \ }a}\Gamma_{\text{ }\mu\rho}^{\nu}h_{b}^{\text{ \ }\rho
}\equiv h_{\nu}^{\text{ \ }a}\nabla_{\mu}h_{b}^{\text{ \ }\nu},
\label{treinta}%
\end{equation}
where $\nabla_{\mu}$ is the usual covariant derivative in the connection
$\Gamma_{\text{ }\mu\rho}^{\nu}$, which acts on external indices only.

\subsection{\textbf{Curvature and Torsion}}

Curvature and torsion are properties of connections \cite{koba}, not of space
itself. This becomes evident if we note that many different connections are
allowed to exist on the very same spacetime \cite{per1}. Of course, when
restricted to the specific case of General Relativity, where only the
Levi-Civita connection is present, universality of gravitation allows it to be
interpreted together with the metric, as part of the spacetime definition.

The curvature and torsion tensors of the connection $A_{\text{ \ }b}^{a}$ are
defined respectively by%
\begin{align}
R_{\text{ \ }b}^{a}  &  =dA_{\text{ \ }b}^{a}+A_{\text{ \ }c}^{a}A_{\text{
\ }b}^{c},\nonumber\\
T^{a}  &  =dh^{a}+A_{\text{ \ }c}^{a}h^{c}. \label{36'}%
\end{align}

This means that the the curvature and torsion tensors for the connection
$A_{\text{ \ }c\nu}^{a}$ are given by%
\begin{align}
R_{\text{ \ }b\nu\mu}^{a} &  =\partial_{\nu}A_{\text{ \ }b\mu}^{a}%
-\partial_{\mu}A_{\text{ \ }b\nu}^{a}+A_{\text{ \ }c\nu}^{a}A_{\text{ \ }b\mu
}^{c}-A_{\text{ \ }c\mu}^{a}A_{\text{ \ }b\nu}^{c},\label{treintaysiete}\\
T_{\text{ }\nu\mu}^{a} &  =\partial_{\nu}h_{\text{ \ }\mu}^{a}-\partial_{\mu
}h_{\text{ \ }\nu}^{a}+A_{\text{ \ }c\nu}^{a}h_{\text{ \ }\mu}^{c}-A_{\text{
\ }c\mu}^{a}h_{\text{ \ }\nu}^{c},\label{treintayocho}%
\end{align}
which can be written also as%
\begin{align}
R_{\text{ \ }\lambda\nu\mu}^{\rho} &  =h_{a}^{\text{ \ }\rho}h_{\text{
\ }\lambda}^{b}R_{\text{ \ }b\nu\mu}^{a}=\partial_{\nu}\Gamma_{\text{
\ }\lambda\mu}^{\rho}-\partial_{\mu}\Gamma_{\text{ \ }\lambda\nu}^{\rho
}+\Gamma_{\text{ \ }c\nu}^{\rho}\Gamma_{\text{ \ }\lambda\mu}^{c}%
-\Gamma_{\text{ \ }c\mu}^{\rho}\Gamma_{\text{ \ }\lambda\nu}^{c},\\
T_{\text{ }\nu\mu}^{\rho} &  =h_{a}^{\text{ \ }\rho}T_{\text{ }\nu\mu}%
^{a}=\Gamma_{\text{ \ }\mu\nu}^{\rho}-\Gamma_{\text{ \ }\nu\mu}^{\rho},\\
T_{\text{ }\nu\mu}^{a} &  =\Gamma_{\text{ \ }\mu\nu}^{a}-\Gamma_{\text{ \ }%
\nu\mu}^{a},
\end{align}

\subsection{Decomposition of the connection}

From metricity condition we have
\begin{equation}
\nabla_{\alpha}g_{\mu\nu}=\partial_{\alpha}g_{\mu\nu}-\Gamma_{\alpha\nu
}^{\beta}g_{\beta\mu}-\Gamma_{\alpha\mu}^{\beta}g_{\beta\nu}=0.\label{dc1}%
\end{equation}
Using the cyclicity of the indices and the identity $\Gamma_{\mu\nu}^{\beta
}=\Gamma_{\left(  \mu\nu\right)  }^{\beta}+\Gamma_{\left[  \mu\nu\right]
}^{\beta}=\Gamma_{\left(  \mu\nu\right)  }^{\beta}+T_{\mu\nu}^{\beta}$, where
$T_{\mu\nu}^{\beta}=\Gamma_{\left[  \mu\nu\right]  }^{\beta}$ is the tensor
torsion, we find that a general linear connection can be decomposed according to%

\begin{align}
\Gamma_{\mu\nu}^{\beta} &  =\Gamma_{\left(  \mu\nu\right)  }^{\beta}+T_{\text{
\ }\mu\nu}^{\beta}=\mathring{\Gamma}_{\mu\nu}^{\beta}-T_{\nu\text{ \ \ }\mu
}^{\text{ \ \ }\beta}-T_{\mu\text{ \ \ \ \ \ }\nu}^{\text{ \ \ \ }\beta
}+T_{\text{ \ }\mu\nu}^{\beta}\nonumber\\
\Gamma_{\mu\nu}^{\text{ \ \ \ }\beta} &  =\mathring{\Gamma}_{\mu\nu}^{\text{
\ \ \ }\beta}+K_{\mu\nu}^{\text{ \ \ \ }\beta},
\end{align}
where $\mathring{\Gamma}_{\mu\nu}^{\beta}$ is the torsionless Christoffel or
levi-Civita connection, and
\begin{equation}
K_{\mu\nu}^{\text{ \ \ \ }\beta}=T_{\mu\nu}^{\text{ \ \ \ }\beta}+T_{\text{
\ \ }\mu\nu}^{\text{ }\beta}-T_{\mu\text{ \ \ \ \ \ }\nu}^{\text{ \ \ \ }%
\beta},
\end{equation}
is the contorsion tensor.

In terms of the spin connection $A_{\text{ \ }b\nu}^{a},$ this decomposition
takes the form
\begin{equation}
A_{\text{ \ }b\nu}^{a}=\mathring{A}_{\text{ \ }b\nu}^{a}+K_{\text{ \ }b\nu
}^{a},\label{10'}%
\end{equation}
where $\mathring{A}_{\text{ \ }b\nu}^{a}$ is the spin connection of general
relativity. On the other hand, by defining the $1$-form connection as
$A_{\text{ \ }b}^{a}=A_{\text{ \ }b\nu}^{a}dx^{\nu},$ we can write%
\begin{equation}
A^{ab}=\mathring{A}^{ab}+K^{ab}.\label{10''}%
\end{equation}

\subsubsection{\textbf{Torsion in anholonomics coordinates}}

From Eqs. (\ref{36'}), (\ref{Ap5}), (\ref{Ap6}), we can see that
\begin{equation}
T^{\text{ }a}=-\frac{1}{2}f_{bc}^{\text{ }a}h^{b}h^{c}+A_{\text{ }b}^{a}h^{b},
\end{equation}
from where%

\begin{equation}
T_{\text{ \ }bc}^{\text{ }a}=A_{\text{ \ }cb}^{a}-A_{\text{ \ }bc}%
^{a}-f_{\text{ \ }bc}^{\text{ }a}.\label{t1}%
\end{equation}
This means that%

\begin{equation}
A_{\text{ \ }bc}^{a}=\frac{1}{2}\left(  T_{b\text{ \ }c}^{\text{ \ }%
a}+f_{b\text{ \ }c}^{\text{ \ }a}+T_{c\text{ \ }b}^{\text{ }a}+f_{c\text{
\ }b}^{\text{ }a}-T_{\text{ \ }bc}^{\text{ }a}-f_{\text{ \ }bc}^{\text{ }%
a}\right)  ,
\end{equation}
or equivalently%
\begin{equation}
A_{\text{ \ }bc}^{a}=\mathring{A}_{\text{ \ }bc}^{a}+K_{\text{ \ }bc}^{a},
\end{equation}
where
\begin{align}
\mathring{A}_{\text{ \ }bc}^{a} &  =\frac{1}{2}\left(  f_{b\text{ \ }%
c}^{\text{ \ }a}+f_{c\text{ \ }b}^{\text{ }a}-f_{\text{ \ }bc}^{\text{ }%
a}\right)  ,\nonumber\\
K_{\text{ \ }bc}^{a} &  =\frac{1}{2}\left(  T_{b\text{ \ }c}^{\text{ \ }%
a}+T_{c\text{ \ }b}^{\text{ }a}-T_{\text{ \ }bc}^{\text{ }a}\right)  .
\end{align}

\subsection{Gravitational Coupling}

From Appendix B we known that the full gravitational coupling prescription is
composed of two parts:

\begin{enumerate}
\item[(1)] the first correspond to the universal, translational coupling
prescription, which is represented by%
\begin{equation}
e_{\text{ \ }\mu}^{a}\partial_{a}\psi\longrightarrow h_{\text{ \ }\mu}%
^{a}\partial_{a}\psi\label{65}%
\end{equation}

\item[(2)] the second correspond to the non-universal spin coupling
prescription%
\begin{equation}
\partial_{a}\psi\longrightarrow\text{ }D_{a}\psi.\label{66}%
\end{equation}

\end{enumerate}

Put together, they yield the full gravitational coupling prescription,
\begin{equation}
e_{\text{ \ }\mu}^{a}\partial_{a}\psi\longrightarrow h_{\text{ \ }\mu}%
^{a}D_{a}\psi\label{67}%
\end{equation}

Equivalently, we can write%
\begin{equation}
\partial_{\mu}\psi\longrightarrow D_{\mu}\psi.\label{68}%
\end{equation}

It is important keep in mind that this is the gravitational coupling procedure
that follows from the general covariance principle, i.e., from the strong
equivalence principle (see Appendix B).

\subsection{Connections}

It is a known fact that the general covariance principle does not determine
uniquely the Lorentz connection $A_{\text{ \ \ }\mu}^{ab}$. In fact, from the
point of view of the coupling procedure, the connection can be chosen freely
among the many possibilities, each one characterized by a connection with
different values of curvature and torsion. Due to the identity (\ref{10'}) we
have that any one of the choices will give rise to a coupling procedure
equivalent to the coupling prescription of General Relativity.

However, there is a strong constraint that must be taken into account.
Considering that the source of gravitation is the symmetric energy-momentum
tensor, which has ten independent components, the gravitational field
equations will be constituted by a set of ten independent differential equations.

The choice of the connection is therefore restricted to not exceed ten
independent components, otherwise the field equations will be unable to
determine it univocally. There are only two choices that respect this constraint.

\textbf{General relativity connection: }The first case involves choosing a
connection without torsion, i.e.,
\begin{equation}
A_{\text{ \ \ }\mu}^{ab}=\mathring{A}_{\text{ \ \ }\mu}^{ab}, \label{71}%
\end{equation}
so the coupling procedure is given by%
\begin{equation}
e_{\text{ \ }\mu}^{a}\partial_{a}\psi\longrightarrow h_{\text{ \ }\mu}%
^{a}\mathring{D}_{a}\psi, \label{72}%
\end{equation}
or equivalently,%
\begin{equation}
\partial_{\mu}\psi\longrightarrow\mathring{D}_{\mu}\psi, \label{73}%
\end{equation}
where $\mathring{D}_{\mu}$ is the covariant derivative corresponding to
connection $\mathring{A}_{\text{ \ \ }\mu}^{ab}.$

Since the connection $\mathring{A}_{\text{ \ \ }\mu}^{ab}$ is completely
determined by the spacetime metric, no additional degrees of freedom is
introduced by this choice. In fact, the linear connection $\mathring{\Gamma
}_{\text{ \ }\nu\mu}^{\rho}$ corresponding to $\mathring{A}_{\text{ \ \ }\mu
}^{ab}$ is just the Christoffel connection of the metric $g_{\mu\nu}$ :%

\begin{equation}
\mathring{\Gamma}_{\text{ \ }\nu\mu}^{\rho}=\frac{1}{2}g^{\rho\lambda}\left(
\partial_{\nu}g_{\lambda\mu}+\partial_{\mu}g_{\lambda\nu}-\partial_{\lambda
}g_{\nu\mu}\right)  .\label{74}%
\end{equation}
The gravitational theory based on such connection is general relativity.

\textbf{Teleparallel connection: }A second possible choice that, like in the
case of general relativity, does not introduce any additional degree of
freedom into the theory, is to assume that the Lorentz connection $A_{\text{
\ \ }\mu}^{ab}$ does not represent gravitation, but only inertial effects.
This means to choose $A_{\text{ \ \ }c\mu}^{b}$ as the inertial connection%

\begin{equation}
\dot{A}_{\text{ \ \ }c\mu}^{b}=\Lambda_{\text{ \ }d}^{b\text{ }}\partial_{\mu
}\Lambda_{c}^{\text{ \ }d\text{ }}, \label{75}%
\end{equation}
where the quantities related to teleparallel gravity will be denoted with a
point $\cdot.$

The gravitational theory corresponding to this choice is just teleparallel
gravity. In this theory, the gravitational field is fully represented by the
the gauge potential $e_{\text{ \ }\mu}^{a}$, which appears as the non-trivial
part of the tetrad field: $h_{\text{ \ }\mu}^{a}=\delta_{\text{ \ }\mu}%
^{a}+e_{\text{ \ }\mu}^{a}$. \ In teleparallel gravity, therefore, Lorentz
connections keep their special relativity role of representing inertial
effects only. \ The teleparallel coupling procedure is then given by
\begin{equation}
e_{\text{ \ }\mu}^{a}\partial_{a}\psi\longrightarrow h_{\text{ \ }\mu}%
^{a}\ddot{D}_{a}\psi.\label{76}%
\end{equation}
Alternatively, one can write
\begin{equation}
\partial_{\mu}\psi\longrightarrow\ddot{D}_{\mu}\psi,\label{77}%
\end{equation}
where $\ddot{D}_{\mu}$ is the covariant derivative corresponding to the
connections $\dot{A}_{\text{ \ \ }c\mu}^{b}-\dot{K}_{\text{ \ \ }c\mu}^{b}.$

Due to identity (\ref{10'}), this coupling procedure is equivalent to the
coupling prescription of general relativity. However, the gravitational theory
based on the spin connection (\ref{75}), although physically equivalent to
general relativity is, conceptually speaking, completely different. In
particular, since that connection represents inertial effects only, the
gravitational field in this theory turns out to be fully represented by the
translational gauge potential $e_{\text{ \ }\mu}^{a}$, as it should be for a
gauge theory for the translation group.

\subsubsection{Curvature versus Torsion}

The curvature of the teleparallel connection (\ref{75}) vanishes identically%
\begin{equation}
\dot{R}^{ab}=d\dot{A}^{ab}+\dot{A}_{\text{ \ \ }c}^{a}\dot{A}^{cb}=0.
\label{81}%
\end{equation}

For \ a tetrad involving a non trivial translational gauge potential $e^{a},$
i.e., for $h^{a}=\dot{D}x^{a}+e^{a},$ the torsion is non-vanishing%
\begin{equation}
\dot{T}^{a}=\dot{D}h^{a}=dh^{a}+\dot{A}_{\text{ \ \ }c}^{a}h^{c}%
\neq0.\label{83}%
\end{equation}

This connection can be considered a kind of \ "dual" of the general relativity
connection, which is a connection with vanishing torsion. In fact, from
(\ref{veintinueve}) we can see that $\mathring{T}_{\text{ \ }\mu\nu}%
^{a}=\Gamma_{\text{ \ }\mu\nu}^{a}-\Gamma_{\text{ \ }\nu\mu}^{a}=\mathring
{D}_{\mu}h_{\nu}^{a}-\mathring{D}_{\nu}h_{\mu}^{a}$, which can be also written
as%
\begin{equation}
\mathring{T}^{a}=\mathring{D}h^{a}=dh^{a}+\mathring{A}_{\text{ \ \ }c}%
^{a}h^{c}=0,\label{84}%
\end{equation}
but no-vanishing curvature
\begin{equation}
\mathring{R}^{ab}=d\mathring{A}^{ab}+\mathring{A}_{\text{ \ \ }c}^{a}\text{
}\mathring{A}^{cb}\neq0.\label{28'}%
\end{equation}

The linear connection corresponding to the spin connection $\dot{A}_{\text{
\ \ }b\nu}^{a}$ is given by
\begin{equation}
\dot{\Gamma}_{\text{ \ }\nu\mu}^{\rho}=h_{a}^{\text{ \ }\rho}\left(
\partial_{\mu}h_{\text{ \ }\nu}^{a}+\dot{A}_{\text{ \ \ }b\mu}^{a}h_{\text{
\ }\nu}^{b}\right)  \equiv h_{a}^{\text{ \ }\rho}\dot{D}_{\mu}h_{\text{ \ }%
\nu}^{a}, \label{86}%
\end{equation}

which is the so-called Weitzenb\"{o}ck connection. Its definition is
equivalent to the identity%

\begin{equation}
\partial_{\mu}h_{\text{ \ }\nu}^{a}+\dot{A}_{\text{ \ \ }b\mu}^{a}h_{\text{
\ }\nu}^{b}-\dot{\Gamma}_{\text{ \ }\nu\mu}^{\rho}h_{\text{ \ }\rho}%
^{a}=0.\label{87}%
\end{equation}

In the class of frames in which the spin connection $\dot{A}_{\text{ \ \ }%
b\mu}^{a}$ vanishes, it becomes
\begin{equation}
\partial_{\mu}h_{\text{ \ }\nu}^{a}-\dot{\Gamma}_{\text{ \ }\nu\mu}^{\rho
}h_{\text{ \ }\rho}^{a}=0,\label{88}%
\end{equation}

which is the absolute or distant parallelism condition from where teleparallel
gravity got its name.

The Weitzenbock connection $\dot{\Gamma}_{\text{ \ }\mu\nu}^{\rho}$ is related
to the Levi-Civita connection $\mathring{\Gamma}_{\text{ \ }\mu\nu}^{\rho}$ of
General Relativity by%

\begin{equation}
\dot{\Gamma}_{\text{ \ }\mu\nu}^{\rho}=\mathring{\Gamma}_{\text{ \ }\mu\nu
}^{\rho}+\dot{K}_{\text{ \ }\mu\nu}^{\rho}. \label{89}%
\end{equation}

In terms of it, the Weitzenb\"{o}ck torsion is written as
\begin{equation}
\dot{T}_{\text{ \ }\mu\nu}^{\rho}=\dot{\Gamma}_{\text{ \ }\nu\mu}^{\rho}%
-\dot{\Gamma}_{\text{ \ }\mu\nu}^{\rho},\label{90}%
\end{equation}

whereas the (vanishing) Weitzenbock curvature is given by%

\begin{equation}
\dot{R}_{\text{ }\rho\nu\mu}^{\lambda}=\partial_{\nu}\dot{\Gamma}_{\text{
\ }\rho\mu}^{\lambda}-\partial_{\mu}\dot{\Gamma}_{\text{ \ \ }\rho\nu
}^{\lambda}+\dot{\Gamma}_{\text{ \ \ }\sigma\nu}^{\lambda}\dot{\Gamma}_{\text{
\ \ }\rho\mu}^{\sigma}-\dot{\Gamma}_{\text{ \ \ }\sigma\mu}^{\lambda}%
\dot{\Gamma}_{\text{ \ }\rho\nu}^{\sigma}=0. \label{91}%
\end{equation}

\section{Lanczos-Lovelock gravity theory}

The simplest action for a gravitational theory satisfying the requirement to
be invariant under local Lorentz transformations and general coordinate
transformations can be written in terms of the vielbein $e^{a}$ and the spin
connection $\omega^{ab}$ which enters through the curvature $R^{ab}$. To find
the mentioned action we remember that the differential forms are invariant
under general transformations of coordinates and under transformations of
Lorentz. So we can postulate that in $n$ dimensions the searched Lagrangian is
given by an $n$-form, which when integrated into the variety leads us to
action,%
\[
S=\int\mathcal{L}_{n},
\]
with%
\[
\mathcal{L}_{n}=\varepsilon_{a_{1}a_{2}\cdot\cdot\cdot\cdot a_{n}}%
R^{a_{1}a_{2}}\cdot\cdot\cdot R^{a_{2p-1}a_{2p}}e^{a_{2p+1}}\cdot\cdot
\cdot\cdot e^{a_{n}}%
\]
where we can see that $\mathcal{L}_{n}$ can be understood or interpreted as a
Euler class dimensionally continued

\subsection{The Euler classes}

Let $M$ be a $2p$-dimensional orientable Riemannian manifold and let $TM$ be
the tangent bundle of $M$. We denote the curvature by $R^{ab}$. It is always
possible to reduce the structure group of $TM$ down to $SO(2p)$ by employing
an orthonormal frame. If $M$ is a $2p$-dimensional surface, then the Euler
class is given by in terms of curvature as%

\[
e_{2p}(M)\sim\varepsilon_{a_{1}a_{2}\cdot\cdot\cdot a_{2p}}R^{a_{1}a_{2}}%
\cdot\cdot\cdot R^{a_{2p-1}a_{2p}}%
\]
where $\varepsilon_{a_{1}a_{2}\cdot\cdot\cdot a_{2p}}$ are invariant tensors
under the Lorentz rotation group $SO(2p).$

If $p=1$, the Euler class is given by%

\begin{equation}
{\Large e}_{2}(M)\sim\varepsilon_{ab}R^{ab}, \label{g1}%
\end{equation}

where $\varepsilon_{ab}$ are invariant tensors under the Lorentz rotation
group $SO(2).$

In a four-dimensional $M$ manifold the Euler class has the form
\begin{equation}
{\Large e}_{4}(M)\sim\varepsilon_{abcd}R^{ab}R^{cd},\label{g2}%
\end{equation}
where $\varepsilon_{abcd}$ are invariant tensors under the Lorentz rotation
group $SO(4).$

In the case that $M$ is a six-dimensional manifold we have that the Euler
class is given by
\begin{equation}
{\Large e}_{4}(M)\sim\varepsilon_{abcdef}R^{ab}R^{cd}R^{ef}.\label{g3}%
\end{equation}

\subsection{\textbf{Lanczos Lovelock action}}

\textbf{Theorem [Lovelock, 1971]: }The most general action for the metric
satisfying the criteria of general covariance and second-order field equations
for $d>4$ is a polynomial of degree $\left[  d/2\right]  $ in the curvature
known as the Lanczos-Lovelock gravity theory $\left(  LL\right)  $
\cite{lanc,lovel}. The $LL$ lagrangian in a $d$-dimensional Riemannian
manifold can be defined as a linear combination of the dimensional
continuation of all the Euler classes of dimension $2p<d$ \cite{zum,teit}:
\begin{equation}
S=\int\sum_{p=0}^{\left[  d/2\right]  }\alpha_{p}L^{(p)},\label{uno}%
\end{equation}
where $\alpha_{p}$ are arbitrary constants and
\begin{equation}
L_{p}=\varepsilon_{a_{1}a_{2}\cdot\cdot\cdot\cdot\cdot\cdot a_{d}}%
R^{a_{1}a_{2}}\cdot\cdot\cdot\cdot R^{a_{2p-1}a_{2p}}e^{a_{2p+1}}\cdot
\cdot\cdot\cdot e^{a_{d}}\label{dos}%
\end{equation}
with $R^{ab}=d\omega^{ab}+\omega_{c}^{a}\omega^{cb}.$ The expression
(\ref{uno}) can be used both for even and for odd dimensions.

The large number of dimensionful constants in the $LL$ theory $\alpha_{p},$
$p=0,1,\cdot\cdot\cdot,\left[  d/2\right]  ,$ which are not fixed from first
principles, contrast with the two constants of the Einstein-Hilbert action.

In Ref. \cite{tron} it was found that these parameters can be fixed in terms
of the gravitational and the cosmological constants, and that the action in
odd dimensions can be formulated as a gauge theory of the $AdS$ group, and in
a particular case, as a gauge theory of the Poincar\'{e} group. This means
that the action is invariant not only under standard local Lorentz rotations
$\delta e^{a}=\kappa_{b}^{a}e^{b};$ $\delta\omega^{ab}=-D\kappa^{ab},$ but
also under local $AdS$ boost $\delta e^{a}=D\rho^{a}$ ; $\delta\omega
^{ab}=\frac{1}{l^{2}}(\rho^{a}e^{b}-\rho^{b}e^{a}),$ where $l$ is a length
parameter. They also show that this situation is not possible in even
dimensions where the action is invariant only under local Lorentz rotations in
the same way as is the Einstein-Hilbert action.

\subsection{\textbf{The local AdS Chern-Simons and Born-Infeld like gravity}}

The $LL$ action is a polynomial of degree $\left[  d/2\right]  $ in curvature,
which can be written in terms of the Riemann curvature and the vielbein
$e^{a}$ in the form (\ref{uno}),(\ref{dos}). In first order formalism, the
$LL$ action is regarded as a functional of the vielbein and spin connection,
and the corresponding field equations obtained by varying with respect to
$e^{a}$ and $\omega^{ab}$ read \cite{tron}%

\begin{equation}
\varepsilon_{a}=\sum_{p=0}^{\left[  \left(  d-1\right)  /2\right]  }\alpha
_{p}(d-2p)\varepsilon_{a}^{p}=0\label{tres}%
\end{equation}%
\begin{equation}
\varepsilon_{ab}=\sum_{p=1}^{\left[  \left(  d-1\right)  /2\right]  }%
\alpha_{p}p(d-2p)\varepsilon_{ab}^{p}=0\label{cuatro}%
\end{equation}
where we have defined
\[
\varepsilon_{a}^{p}=\varepsilon_{ab_{1}b_{2}\cdot\cdot\cdot\cdot\cdot\cdot
b_{d-1}}R^{b_{1}b_{2}}\cdot\cdot\cdot
\]%
\begin{equation}
\cdot\cdot\cdot R^{b_{2p-1}b_{2p}}e^{b_{2p+1}}\cdot\cdot\cdot\cdot e^{b_{d-1}%
}\label{cinco}%
\end{equation}%
\[
\varepsilon_{ab}^{p}=\varepsilon_{aba_{3}\cdot\cdot\cdot\cdot\cdot\cdot a_{d}%
}R^{a_{3}a_{4}}\cdot\cdot\cdot
\]%
\begin{equation}
\cdot\cdot\cdot\cdot R^{a_{2p-1}a_{2p}}T^{a_{2p+1}}e^{a_{2p+2}}\cdot\cdot
\cdot\cdot e^{a_{d}}.\label{seis}%
\end{equation}
Here $T^{a}=de^{a}+\omega_{\text{ \ }b}^{a}e^{b}$ is the torsion $2$-form.
Using the Bianchi identity one finds \cite{tron}
\begin{equation}
D\varepsilon_{a}=\sum_{p=1}^{\left[  \left(  d-1\right)  /2\right]  }%
\alpha_{p-1}(d-2p+2)(d-2p+1)e^{b}\varepsilon_{ba}^{p}.\label{siete}%
\end{equation}
Moreover
\begin{equation}
e^{b}\varepsilon_{ba}=\sum_{p=1}^{\left[  \left(  d-1\right)  /2\right]
}\alpha_{p}p(d-2p)e^{b}\varepsilon_{ba}^{p}.\label{ocho}%
\end{equation}
From (\ref{siete}) and (\ref{ocho}) one finds for $d=2n-1$%
\[
\alpha_{p}=\alpha_{0}\frac{(2n-1)(2\gamma)^{p}}{(2n-2p-1)}\left(
\genfrac{}{}{0pt}{}{n-1}{p}%
\right)  ,\qquad\alpha_{0}=\frac{\kappa}{dl^{d-1}},\qquad
\]%
\begin{equation}
\gamma=-sign(\Lambda)\frac{l^{2}}{2},\label{nueve}%
\end{equation}
where for any dimensions, $l$ is a length parameter related to the
cosmological constant by $\Lambda=\pm(d-1)(d-2)/2l^{2}.$ With these
coefficients, the $LL$ action is invariant under local Lorentz rotations and
under local AdS boosts.

For $d=2n$ it is necessary to write equation (\ref{siete}) in the form
\cite{tron}
\begin{align}
D\varepsilon_{a}  & =T^{a}\sum_{p=1}^{\left[  n-1\right]  }2\alpha
_{p-1}(n-p+1)\mathcal{T}_{ab}^{p}\nonumber\\
& -\sum_{p=1}^{\left[  n-1\right]  }4\alpha_{p-1}(n-p+1)(n-p)e^{b}%
\varepsilon_{ba}^{p}\label{diez}%
\end{align}
with
\begin{equation}
\mathcal{T}_{ab}=\frac{\delta L}{\delta R^{ab}}=\sum_{p=1}^{\left[  \left(
d-1\right)  /2\right]  }\alpha_{p}p\mathcal{T}_{ab}^{p},\label{once}%
\end{equation}
where
\[
\mathcal{T}_{ab}^{p}=\varepsilon_{aba_{3}\cdot\cdot\cdot\cdot a_{d}}%
R^{a_{3}a_{4}}\cdot\cdot\cdot
\]%
\begin{equation}
\cdot\cdot\cdot R^{a_{2p-1}a_{2p}}T^{a_{2p+1}}e^{a_{2p+2}}\cdot\cdot\cdot
e^{a_{d}}.\label{doce}%
\end{equation}

The comparison between (\ref{ocho}) and (\ref{diez}) leads to \cite{tron}
\begin{equation}
\alpha_{p}=\alpha_{0}(2\gamma)^{p}\binom{n}{p}.\label{trece}%
\end{equation}
With these coefficients the $LL$ action, in the same way as the
Einstein-Hilbert action, is invariant only under local Lorentz rotations.

\subsection{\textbf{Theories described by a generalized action}}

In Ref. \cite{criso} was found a class of gravitational theories described by
the action
\begin{equation}
S=\int\sum\alpha_{p}L^{(p)}=\kappa\int\sum_{p=0}^{k}C_{p}^{k}L^{(p)}%
,\label{d1}%
\end{equation}
where
\begin{equation}
\alpha_{p}=\kappa C_{p}^{k}=\left\{
\begin{array}
[c]{c}%
\kappa\frac{\alpha_{0}d(2\gamma)^{p}}{(d-2p)}\binom{k}{p},\quad p\leq k,\\
0,\qquad p>k,
\end{array}
\right.  \label{d1'}%
\end{equation}
with $1\leq k\acute{\leq}[(d-1)/2]$ and where $L_{p}$ is given by
\begin{equation}
L_{p}=\varepsilon_{a_{1}a_{2}\cdot\cdot\cdot\cdot\cdot\cdot a_{d}}%
R^{a_{1}a_{2}}\cdot\cdot\cdot\cdot R^{a_{2p-1}a_{2p}}e^{a_{2p+1}}\cdot
\cdot\cdot\cdot e^{a_{d}}\label{d2}%
\end{equation}
where $R^{ab}=d\omega^{ab}+\omega_{\text{ \ }c}^{a}\omega^{cb}$ is the
curvarure $2$-form and $e^{a}$ is the vielbein $1$-form. For a given dimension
$d$, the $\alpha_{p}$ coefficients give rise to a family of inequivalent
theories, labeled by the integer $k\in\left\{  1\cdot\cdot\cdot\cdot
\lbrack(d-1)/2]\right\}  $ which represents the highest power of curvature in
the lagrangian. This set of theories possees only two fundamental constants,
$\kappa$ and $l$, related to the gravitational constant $G$ and the
cosmological constant $\Lambda$ \cite{tron}. For $k=1$, the Einstein-Hilbert
action is recovered, while for the largest values of $k$, that is
$k=[(d-1)/2],$ Born-Infeld and Chern-Simons theories are obtained. These three
cases exhaust the different possibilities up to six dimensions, and new
interesting cases arise for $d\geq7$. For instance, the case with $k=2$, which
is described by the action \cite{criso}
\[
S_{2}=\kappa\int\varepsilon_{a_{1}a_{2}\cdot\cdot\cdot\cdot a_{d}}%
(\frac{l^{-4}}{d}e^{a_{1}}\cdot\cdot\cdot e^{a_{d}}+\frac{2l^{-2}}%
{d-2}R^{a_{1}a_{2}}e^{a_{3}}\cdot\cdot\cdot e^{a_{d}}%
\]%
\begin{equation}
+\frac{1}{d-4}R^{a_{1}a_{2}}R^{a_{3}a_{4}}e^{a_{5}}\cdot\cdot\cdot e^{a_{d}%
})\label{a1}%
\end{equation}
exists only for $d>4:$ in five dimensions this theory is equivalent to
Chern-Simons; for $d=6$ it is equivalent to Born-Infeld and for $d=7$ and up,
if defines a new class of theories.

At the end of the range, $k=[(d-1)/2]$, even and odd dimensions must be
distinguished. When $d=2n-1$, the maximum value of $k$ is $n-1$, and the
corresponding lagrangian of the action
\[
S_{3}=\int\sum_{p=0}^{\left[  d/2\right]  }\frac{\kappa}{d-2p}\binom{n-1}%
{p}l^{2p-d+1}\varepsilon_{a_{1}a_{2}\cdot\cdot\cdot\cdot\cdot\cdot a_{d}%
}R^{a_{1}a_{2}}\cdot\cdot\cdot
\]%
\begin{equation}
\cdot\cdot\cdot\cdot R^{a_{2p-1}a_{2p}}e^{a_{2p+1}}\cdot\cdot\cdot\cdot
e^{a_{d}}\label{a2}%
\end{equation}
is a Chern-Simons $\left(  2n-1\right)  $-form. This action \cite{tron} is
invariant not only under standard local Lorentz rotations
\begin{equation}
\delta e^{a}=\kappa_{\text{ }b}^{a}e^{b},\quad\delta\omega^{ab}=-D\kappa
^{ab},\label{d3}%
\end{equation}
but also under a local $AdS$ boost
\begin{equation}
\delta e^{a}=D\rho^{a},\quad\delta\omega^{ab}=\frac{1}{l^{2}}\left(  \rho
^{a}e^{b}-\rho^{b}e^{a}\right)  .\label{d4}%
\end{equation}

For $d=2n$ and $k=n-1$, the action is given by
\[
S=\int\sum_{p=0}^{\left[  d/2\right]  }\frac{\kappa}{2n}\binom{n}{p}%
l^{2p-d+1}\varepsilon_{a_{1}a_{2}\cdot\cdot\cdot\cdot\cdot\cdot a_{d}}%
R^{a_{1}a_{2}}\cdot\cdot\cdot
\]%
\begin{equation}
\cdot\cdot\cdot R^{a_{2p-1}a_{2p}}e^{a_{2p+1}}\cdot\cdot\cdot\cdot e^{a_{d}}.
\label{a3}%
\end{equation}
This action \cite{tron} is invariant under standard local Lorentz rotations
but it is not invariant under local $AdS$ boosts.

\section{\textbf{Teleparallel Equivalent of Euler Classes}}

In this section we will consider the teleparallel equivalent of Euler classes
in different dimensions. From (\ref{10''}), (\ref{81}), (\ref{83}),
(\ref{84}), (\ref{28'}) we can see that introducing (\ref{10''}) into
(\ref{81}) we find%
\begin{equation}
\dot{R}^{ab}=\left(  d\mathring{A}^{ab}+\mathring{A}_{\text{ \ }c}%
^{a}\mathring{A}^{cb}\right)  +\left(  d\dot{K}^{ab}+\mathring{A}_{\text{
\ }c}^{a}\dot{K}^{cb}+\dot{K}_{\text{ \ }c}^{a}\mathring{A}^{cb}\right)
+\dot{K}_{\text{ \ }c}^{a}\dot{K}^{cb},
\end{equation}
so that%
\begin{equation}
\dot{R}^{ab}=\mathring{R}^{ab}+\mathring{D}\dot{K}^{ab}+\dot{K}_{\text{ \ }%
c}^{a}\dot{K}^{cb}.\label{86'}%
\end{equation}

On the other hand, introducing (\ref{10'}) into (\ref{86'}) we have%
\[
\mathring{R}_{\text{ }b}^{a}=\left(  d\dot{A}^{ab}+\dot{A}_{\text{ \ }c}%
^{a}\dot{A}^{cb}\right)  -\left(  d\dot{K}^{ab}+\dot{A}_{\text{ \ }c}^{a}%
\dot{K}^{cb}+\dot{K}_{\text{ \ }c}^{a}\dot{A}^{cb}\right)  +\dot{K}_{\text{
\ }c}^{a}\dot{K}^{cb},
\]
that is%
\begin{equation}
\dot{R}^{ab}=\mathring{R}^{ab}+\dot{D}\dot{K}^{ab}-\dot{K}_{\text{ \ }c}%
^{a}\dot{K}^{cb}. \label{88'}%
\end{equation}

The equations (\ref{86'}) (\ref{88'}) will be used to obtain the teleparallel
equivalent of the so-called Euler classes. For this we must remember that the
curvature corresponding to the connection $\dot{A}^{ab}$ is zero. From
(\ref{86'}) we see that%

\begin{align}
\mathring{R}^{ab} &  =-\mathring{D}\dot{K}^{ab}-\dot{K}_{\text{ \ }c}^{a}%
\dot{K}^{cb}\label{96''}\\
&  =-\dot{D}\dot{K}^{ab}+\dot{K}_{\text{ \ }c}^{a}\dot{K}^{cb}.
\end{align}

We know that if $M$ is a $2n$-dimensional surface, then the Euler class for
the Levi-Civita connection is given by%

\begin{equation}
e(M)\sim\varepsilon_{a_{1}a_{2}\cdot\cdot\cdot\cdot a_{2n}}\mathring{R}%
^{a_{1}a_{2}}\mathring{R}^{a_{3}a_{4}}\cdot\cdot\cdot\cdot\mathring
{R}^{a_{2n-1}a_{2n}}\label{eu1}%
\end{equation}
From (\ref{96''}), we have%
\begin{align}
e_{2n}(M) &  \sim\varepsilon_{a_{1}a_{2}\cdot\cdot\cdot\cdot a_{2n}}\left(
-1\right)  ^{n}\left(  \dot{K}_{\text{ \ }c}^{a_{1}}\dot{K}^{ca_{2}}%
+\mathring{D}\dot{K}^{a_{1}a_{2}}\right)  \left(  \dot{K}_{\text{ \ }c}%
^{a_{3}}\dot{K}^{ca_{4}}+\mathring{D}\dot{K}^{a_{3}a_{4}}\right)  \cdot
\cdot\cdot\nonumber\\
&  \cdot\cdot\cdot\left(  \dot{K}_{\text{ \ \ \ \ }c_{2n-1}}^{a_{2n-1}}\dot
{K}^{c_{2n-1}a_{2n}}+\mathring{D}\dot{K}^{a_{2n-1}a_{2n}}\right)  ,\label{eu2}%
\end{align}
which can be written in abbreviated notation in the following way:
$\varepsilon_{ab}\left(  \dot{K}_{\text{ \ }c}^{a}\dot{K}^{cb}+\mathring
{D}\dot{K}^{ab}\right)  =\dot{K}\dot{K}+\mathring{D}\dot{K}.$ So that%
\begin{align}
e_{2n}(M) &  \sim\left(  -1\right)  ^{n}\left(  \dot{K}\dot{K}+\mathring
{D}\dot{K}\right)  \left(  \dot{K}\dot{K}+\mathring{D}\dot{K}\right)
\cdot\cdot\cdot\cdot\left(  \dot{K}\dot{K}+\mathring{D}\dot{K}\right)
\nonumber\\
&  =\left(  -1\right)  ^{n}\left(  \dot{K}\dot{K}+\mathring{D}\dot{K}\right)
^{n}.\label{eu3}%
\end{align}
Using the binomial theorem%
\[
\left(  X+Y\right)  ^{n}=\sum_{j}\binom{n}{j}X^{\text{ }n-j}\wedge Y^{j}%
\]
we can write%
\begin{align}
e_{2n}(M) &  \sim\left(  -1\right)  ^{n}\left(  \dot{K}\dot{K}+\mathring
{D}\dot{K}\right)  ^{n}\nonumber\\
&  =\sum_{j=0}^{n}\left(  -1\right)  ^{n}\binom{n}{j}\left(  \dot{K}\dot
{K}\right)  ^{\text{ }n-j}\wedge\left(  \mathring{D}\dot{K}\right)
^{j},\label{eu4}%
\end{align}
where $\dot{K}\dot{K}\equiv\dot{K}_{\text{ \ }c}^{a}\dot{K}^{cb}$ and
$\mathring{D}\dot{K}\equiv\mathring{D}\dot{K}^{ab}.$

Consider now some particular cases .

\subsection{\textbf{Euler Class in two dimensions}}

This case corresponds to the  $n=1$ case of (\ref{eu4}). In fact, for $n=1$,
we have%

\begin{align}
e_{2}(M) &  \sim\left(  -1\right)  ^{1}\binom{1}{0}\left(  \dot{K}\dot
{K}\right)  ^{\text{ }1}\wedge\left(  \mathring{D}\dot{K}\right)  ^{0}+\left(
-1\right)  ^{1}\binom{1}{1}\left(  \dot{K}\dot{K}\right)  ^{\text{ }1-1}%
\wedge\left(  \mathring{D}\dot{K}\right)  ^{1}\nonumber\\
&  =-\left(  \dot{K}\dot{K}\right)  -\left(  \mathring{D}\dot{K}\right)
\nonumber\\
&  =-\varepsilon_{a_{1}a_{2}}\left[  d\dot{K}^{a_{1}a_{2}}+\left(  \dot{K}%
\dot{K}\right)  ^{a_{1}a_{2}}\right]  =-\varepsilon_{a_{1}a_{2}}\Gamma
^{a_{1}a_{2}},\label{eu6}%
\end{align}
where $\Gamma^{a_{1}a_{2}}=d\dot{K}^{a_{1}a_{2}}+\left(  \dot{K}\dot
{K}\right)  ^{a_{1}a_{2}}.$ So that, we can write%
\begin{equation}
e_{2}(M)=-\left\langle \Gamma\right\rangle =-\varepsilon_{a_{1}a_{2}}%
\Gamma^{a_{1}a_{2}}=-\left\langle d\dot{K}+\dot{K}\dot{K}\right\rangle
\label{eu7}%
\end{equation}

\subsection{\textbf{Euler Class in four dimensions}}

We are now in the case $n=2$ of (\ref{eu4})$.$ In fact, for $n=2$, we have
\begin{align}
e_{4}(M) &  \sim\sum_{j=0}^{2}\left(  -1\right)  ^{2}\binom{2}{j}\left(
\dot{K}\dot{K}\right)  ^{\text{ }2-j}\wedge\left(  \mathring{D}\dot{K}\right)
^{j}\nonumber\\
&  =\binom{2}{0}\left(  \dot{K}\dot{K}\right)  ^{\text{ }2-0}\wedge\left(
\mathring{D}\dot{K}\right)  ^{0}+\binom{2}{1}\left(  \dot{K}\dot{K}\right)
^{\text{ }2-1}\wedge\left(  \mathring{D}\dot{K}\right)  ^{1}\nonumber\\
&  +\binom{2}{2}\left(  \dot{K}\dot{K}\right)  ^{\text{ }2-2}\wedge\left(
\mathring{D}\dot{K}\right)  ^{2}\nonumber\\
&  =\varepsilon_{a_{1}a_{2}a_{3}a_{4}}\left(  \mathring{D}\dot{K}^{a_{1}a_{2}%
}\mathring{D}\dot{K}^{a_{3}a_{4}}+2\mathring{D}\dot{K}^{a_{1}a_{2}}\dot
{K}_{\text{ \ }c_{3}}^{a_{3}}\dot{K}^{c_{3}a_{4}}+\dot{K}_{\text{ \ \ }c_{1}%
}^{a_{1}}\dot{K}^{c_{1}a_{2}}\dot{K}_{\text{ \ \ }c_{3}}^{a_{3}}\dot{K}%
^{c_{3}a_{4}}\right)  .\label{eu9}%
\end{align}
Since%
\begin{align*}
\varepsilon_{a_{1}a_{2}a_{3}a_{4}}\mathring{D}\dot{K}^{a_{1}a_{2}}\mathring
{D}\dot{K}^{a_{3}a_{4}} &  =\mathring{D}\left(  \varepsilon_{a_{1}a_{2}%
a_{3}a_{4}}\dot{K}^{a_{1}a_{2}}\mathring{D}\dot{K}^{a_{3}a_{4}}\right)
+\varepsilon_{a_{1}a_{2}a_{3}a_{4}}\dot{K}^{a_{1}a_{2}}\mathring{D}^{2}\dot
{K}^{a_{3}a_{4}}\\
\mathring{D}^{2}\dot{K}^{a_{3}a_{4}} &  =\mathring{R}_{\text{ \ }c_{3}}%
^{a_{3}}\dot{K}^{c_{3}a_{4}}-\mathring{R}_{\text{ \ }c_{4}}^{a_{4}}\dot
{K}^{c_{4}a_{3}},
\end{align*}
we find%
\[
\varepsilon_{a_{1}a_{2}a_{3}a_{4}}\mathring{D}\dot{K}^{a_{1}a_{2}}\mathring
{D}\dot{K}^{a_{3}a_{4}}=\mathring{D}\left(  \varepsilon_{a_{1}a_{2}a_{3}a_{4}%
}\dot{K}^{a_{1}a_{2}}\mathring{D}\dot{K}^{a_{3}a_{4}}\right)  +2\varepsilon
_{a_{1}a_{2}a_{3}a_{4}}\dot{K}^{a_{1}a_{2}}\mathring{R}_{\text{ \ }c_{3}%
}^{a_{3}}\dot{K}^{a_{3}a_{4}},
\]
so that
\begin{align*}
e_{4}(M) &  \sim\varepsilon_{a_{1}a_{2}a_{3}a_{4}}\mathring{R}^{a_{1}a_{2}%
}\mathring{R}^{a_{3}a_{4}}=d\left(  \varepsilon_{a_{1}a_{2}a_{3}a_{4}}\dot
{K}^{a_{1}a_{2}}\mathring{D}\dot{K}^{a_{3}a_{4}}\right)  +2\varepsilon
_{a_{1}a_{2}a_{3}a_{4}}\dot{K}^{a_{1}a_{2}}\mathring{R}_{\text{ \ }c_{3}%
}^{a_{3}}\dot{K}^{c_{3}a_{4}}\\
&  +\varepsilon_{a_{1}a_{2}a_{3}a_{4}}\left(  2\mathring{D}\dot{K}^{a_{1}%
a_{2}}\dot{K}_{\text{ \ }c_{3}}^{a_{3}}\dot{K}^{c_{3}a_{4}}+\dot{K}_{\text{
\ \ }c_{1}}^{a_{1}}\dot{K}^{c_{1}a_{2}}\dot{K}_{\text{ \ \ }c_{3}}^{a_{3}}%
\dot{K}^{c_{3}a_{4}}\right)  ,
\end{align*}
expression that can be written as,
\[
\varepsilon_{a_{1}a_{2}a_{3}a_{4}}\left(  2\mathring{D}\dot{K}^{a_{1}a_{2}%
}\dot{K}_{\text{ \ }c_{3}}^{a_{3}}\dot{K}^{c_{3}a_{4}}\right)  =\mathring
{D}\left(  2\varepsilon_{a_{1}a_{2}a_{3}a_{4}}\dot{K}^{a_{1}a_{2}}\dot
{K}_{\text{ \ }c_{3}}^{a_{3}}\dot{K}^{c_{3}a_{4}}\right)  +4\varepsilon
_{a_{1}a_{2}a_{3}a_{4}}\dot{K}^{a_{1}a_{2}}\mathring{D}\dot{K}_{\text{
\ }c_{3}}^{a_{3}}\dot{K}^{c_{3}a_{4}},
\]
where we have used
\[
\mathring{D}\left(  2\dot{K}^{a_{1}a_{2}}\dot{K}_{\text{ \ }c_{3}}^{a_{3}}%
\dot{K}^{c_{3}a_{4}}\right)  =2\mathring{D}\dot{K}^{a_{1}a_{2}}\dot{K}_{\text{
\ }c_{3}}^{a_{3}}\dot{K}^{c_{3}a_{4}}-2\dot{K}^{a_{1}a_{2}}\mathring{D}\dot
{K}_{\text{ \ }c_{3}}^{a_{3}}\dot{K}^{c_{3}a_{4}}+2\dot{K}^{a_{1}a_{2}}\dot
{K}_{\text{ \ }c_{3}}^{a_{3}}\mathring{D}\dot{K}^{a_{3}a_{4}}.
\]
This means%

\begin{align*}
e_{4}(M) &  \sim\varepsilon_{a_{1}a_{2}a_{3}a_{4}}\mathring{R}^{a_{1}a_{2}%
}\mathring{R}^{a_{3}a_{4}}=d\left(  \varepsilon_{a_{1}a_{2}a_{3}a_{4}}\dot
{K}^{a_{1}a_{2}}\mathring{D}\dot{K}^{a_{3}a_{4}}+2\varepsilon_{a_{1}a_{2}%
a_{3}a_{4}}\dot{K}^{a_{1}a_{2}}\dot{K}_{\text{ \ }c_{3}}^{a_{3}}\dot{K}%
^{c_{3}a_{4}}\right)  \\
&  +2\varepsilon_{a_{1}a_{2}a_{3}a_{4}}\dot{K}^{a_{1}a_{2}}\mathring
{R}_{\text{ \ }c_{3}}^{a_{3}}\dot{K}^{c_{3}a_{4}}+4\varepsilon_{a_{1}%
a_{2}a_{3}a_{4}}\dot{K}^{a_{1}a_{2}}\mathring{D}\dot{K}_{\text{ \ }c_{3}%
}^{a_{3}}\dot{K}^{c_{3}a_{4}}\\
&  +\varepsilon_{a_{1}a_{2}a_{3}a_{4}}\left(  \dot{K}_{\text{ \ \ }c_{1}%
}^{a_{1}}\dot{K}^{c_{1}a_{2}}\dot{K}_{\text{ \ \ }c_{3}}^{a_{3}}\dot{K}%
^{c_{3}a_{4}}\right)  .
\end{align*}
Given that%
\begin{align*}
\mathring{D}\dot{K}^{a_{3}a_{4}} &  =\dot{D}\dot{K}^{a_{3}a_{4}}-2\dot
{K}_{\text{ \ }c_{3}}^{a_{3}}\dot{K}^{c_{3}a_{4}},\\
\mathring{R}^{a_{1}a_{2}} &  =-\dot{D}\dot{K}^{a_{1}a_{2}}+\dot{K}_{\text{
\ \ }c_{1}}^{a_{1}}\dot{K}^{c_{1}a_{2}},
\end{align*}
we can write
\begin{align*}
&  \varepsilon_{a_{1}a_{2}a_{3}a_{4}}\dot{K}^{a_{1}a_{2}}\mathring{D}\dot
{K}^{a_{3}a_{4}}+2\varepsilon_{a_{1}a_{2}a_{3}a_{4}}\dot{K}^{a_{1}a_{2}}%
\dot{K}_{\text{ \ }c_{3}}^{a_{3}}\dot{K}^{c_{3}a_{4}}\\
&  =\varepsilon_{a_{1}a_{2}a_{3}a_{4}}\dot{K}^{a_{1}a_{2}}\dot{D}\dot
{K}^{a_{3}a_{4}},
\end{align*}
and%
\begin{align*}
&  2\varepsilon_{a_{1}a_{2}a_{3}a_{4}}\dot{K}^{a_{1}a_{2}}\mathring{R}_{\text{
\ }c_{3}}^{a_{3}}\dot{K}^{c_{3}a_{4}}+4\varepsilon_{a_{1}a_{2}a_{3}a_{4}}%
\dot{K}^{a_{1}a_{2}}\mathring{D}\dot{K}_{\text{ \ }c_{3}}^{a_{3}}\dot
{K}^{a_{3}a_{4}}\\
&  =2\varepsilon_{a_{1}a_{2}a_{3}a_{4}}\dot{K}^{a_{1}a_{2}}\left(  -\dot
{D}\dot{K}_{\text{ \ }c_{3}}^{a_{3}}+\dot{K}_{\text{ \ \ }d_{3}}^{a_{3}}%
\dot{K}_{\text{ \ }c_{3}}^{d_{3}}\right)  \dot{K}^{c_{3}a_{4}}\\
&  +4\varepsilon_{a_{1}a_{2}a_{3}a_{4}}\dot{K}^{a_{1}a_{2}}\left(  \dot{D}%
\dot{K}_{\text{ \ }c_{3}}^{a_{3}}-2\dot{K}_{\text{ \ \ }d_{3}}^{a_{3}}\dot
{K}_{\text{ \ }c_{3}}^{d_{3}}\right)  \dot{K}^{c_{3}a_{4}}\\
&  =2\varepsilon_{a_{1}a_{2}a_{3}a_{4}}\dot{K}^{a_{1}a_{2}}\dot{D}\dot
{K}_{\text{ \ }c_{3}}^{a_{3}}\dot{K}^{c_{3}a_{4}},
\end{align*}
where $\varepsilon_{a_{1}a_{2}a_{3}a_{4}}\dot{K}^{a_{1}a_{2}}\dot{K}_{\text{
\ \ }d_{3}}^{a_{3}}\dot{K}_{\text{ \ }c_{3}}^{d_{3}}\dot{K}^{c_{3}a_{4}}=0.$
Therefore,%
\begin{align}
e_{4}(M) &  \sim\varepsilon_{a_{1}a_{2}a_{3}a_{4}}\mathring{R}^{a_{1}a_{2}%
}\mathring{R}^{a_{3}a_{4}}\nonumber\\
&  =d\left(  \varepsilon_{a_{1}a_{2}a_{3}a_{4}}\dot{K}^{a_{1}a_{2}}\dot{D}%
\dot{K}^{a_{3}a_{4}}\right)  +2\varepsilon_{a_{1}a_{2}a_{3}a_{4}}\dot
{K}^{a_{1}a_{2}}\dot{D}\dot{K}_{\text{ \ }c_{3}}^{a_{3}}\dot{K}^{c_{3}a_{4}%
}\nonumber\\
&  +\varepsilon_{a_{1}a_{2}a_{3}a_{4}}\dot{K}_{\text{ \ \ }c_{1}}^{a_{1}}%
\dot{K}^{c_{1}a_{2}}\dot{K}_{\text{ \ \ }c_{3}}^{a_{3}}\dot{K}^{c_{3}a_{4}%
}.\label{eu10}%
\end{align}

\subsection{\textbf{Euler Class in six dimensions}}

This case corresponds to the $n=3$ case  of (\ref{eu4}). In fact, for $n=3$,
we have%

\begin{align*}
e_{6}(M) &  \sim\sum_{j=0}^{3}\left(  -1\right)  ^{3}\binom{3}{j}\left(
\dot{K}\dot{K}\right)  ^{\text{ }3-j}\wedge\left(  \mathring{D}\dot{K}\right)
^{j}\\
&  =-\binom{3}{0}\left(  \dot{K}\dot{K}\right)  ^{\text{ }3-0}\wedge\left(
\mathring{D}\dot{K}\right)  ^{0}-\binom{3}{1}\left(  \dot{K}\dot{K}\right)
^{\text{ }3-1}\wedge\left(  \mathring{D}\dot{K}\right)  ^{1}\\
&  -\binom{3}{2}\left(  \dot{K}\dot{K}\right)  ^{\text{ }3-2}\wedge\left(
\mathring{D}\dot{K}\right)  ^{2}-\binom{3}{3}\left(  \dot{K}\dot{K}\right)
^{\text{ }3-3}\wedge\left(  \mathring{D}\dot{K}\right)  ^{3}\\
&  =-\varepsilon_{a_{1}a_{2}a_{3}a_{4}a_{5}a_{6}}\dot{K}_{\text{ \ \ }c_{1}%
}^{a_{1}}\dot{K}^{c_{1}a_{2}}\dot{K}_{\text{ \ \ }c_{3}}^{a_{3}}\dot{K}%
^{c_{3}a_{4}}\dot{K}_{\text{ \ \ }c_{5}}^{a_{5}}\dot{K}^{c_{5}a_{6}%
}-3\varepsilon_{a_{1}a_{2}a_{3}a_{4}a_{5}a_{6}}\dot{K}_{\text{ \ \ }c_{1}%
}^{a_{1}}\dot{K}^{c_{1}a_{2}}\dot{K}_{\text{ \ \ }c_{3}}^{a_{3}}\dot{K}%
^{c_{3}a_{4}}\mathring{D}\dot{K}^{a_{5}a_{6}}\\
&  -3\varepsilon_{a_{1}a_{2}a_{3}a_{4}a_{5}a_{6}}\dot{K}_{\text{ \ \ }c_{1}%
}^{a_{1}}\dot{K}^{c_{1}a_{2}}\mathring{D}\dot{K}^{a_{3}a_{4}}\mathring{D}%
\dot{K}^{a_{5}a_{6}}-\varepsilon_{a_{1}a_{2}a_{3}a_{4}a_{5}a_{6}}\mathring
{D}\dot{K}^{a_{1}a_{2}}\mathring{D}\dot{K}^{a_{3}a_{4}}\mathring{D}\dot
{K}^{a_{5}a_{6}}.
\end{align*}

\subsection{\textbf{Euler Class in eight dimensions}}

If $n=4$ in (\ref{eu4}) we obtain%

\begin{align*}
e_{8}(M)  &  \sim\sum_{j=0}^{4}\left(  -1\right)  ^{4}\binom{4}{j}\left(
\dot{K}\dot{K}\right)  ^{\text{ }4-j}\wedge\left(  \mathring{D}\dot{K}\right)
^{j}\\
&  =\binom{4}{0}\left(  \dot{K}\dot{K}\right)  ^{\text{ }4-0}\wedge\left(
\mathring{D}\dot{K}\right)  ^{0}+\binom{4}{1}\left(  \dot{K}\dot{K}\right)
^{\text{ }4-1}\wedge\left(  \mathring{D}\dot{K}\right)  ^{1}\\
&  +\binom{4}{2}\left(  \dot{K}\dot{K}\right)  ^{\text{ }4-2}\wedge\left(
\mathring{D}\dot{K}\right)  ^{2}+\binom{4}{3}\left(  \dot{K}\dot{K}\right)
^{\text{ }4-3}\wedge\left(  \mathring{D}\dot{K}\right)  ^{3}\\
&  +\binom{4}{4}\left(  \dot{K}\dot{K}\right)  ^{\text{ }4-4}\wedge\left(
\mathring{D}\dot{K}\right)  ^{4}\\
&  =\varepsilon_{a_{1}a_{2}a_{3}a_{4}a_{5}a_{6}a_{7}a_{8}}\dot{K}_{\text{
\ \ }c_{1}}^{a_{1}}\dot{K}^{c_{1}a_{2}}\dot{K}_{\text{ \ \ }c_{3}}^{a_{3}}%
\dot{K}^{c_{3}a_{4}}\dot{K}_{\text{ \ \ }c_{5}}^{a_{5}}\dot{K}^{c_{5}a_{6}%
}\dot{K}_{\text{ \ \ }c_{7}}^{a_{7}}\dot{K}^{c_{7}a_{8}}\\
&  +4\varepsilon_{a_{1}a_{2}a_{3}a_{4}a_{5}a_{6}a_{7}a_{8}}\dot{K}_{\text{
\ \ }c_{1}}^{a_{1}}\dot{K}^{c_{1}a_{2}}\dot{K}_{\text{ \ \ }c_{3}}^{a_{3}}%
\dot{K}^{c_{3}a_{4}}\dot{K}_{\text{ \ \ }c_{5}}^{a_{5}}\dot{K}^{c_{5}a_{6}%
}\mathring{D}\dot{K}^{a_{7}a_{8}}\\
&  +6\varepsilon_{a_{1}a_{2}a_{3}a_{4}a_{5}a_{6}a_{7}a_{8}}\dot{K}_{\text{
\ \ }c_{1}}^{a_{1}}\dot{K}^{c_{1}a_{2}}\dot{K}_{\text{ \ \ }c_{3}}^{a_{3}}%
\dot{K}^{c_{3}a_{4}}\mathring{D}\dot{K}^{a_{5}a_{6}}\mathring{D}\dot{K}%
^{a_{7}a_{8}}\\
&  +4\varepsilon_{a_{1}a_{2}a_{3}a_{4}a_{5}a_{6}a_{7}a_{8}}\dot{K}_{\text{
\ \ }c_{1}}^{a_{1}}\dot{K}^{c_{1}a_{2}}\mathring{D}\dot{K}^{a_{3}a_{4}%
}\mathring{D}\dot{K}^{a_{5}a_{6}}\mathring{D}\dot{K}^{a_{7}a_{8}}\\
&  +\varepsilon_{a_{1}a_{2}a_{3}a_{4}a_{5}a_{6}a_{7}a_{8}}\mathring{D}\dot
{K}^{a_{1}a_{2}}\mathring{D}\dot{K}^{a_{3}a_{4}}\mathring{D}\dot{K}%
^{a_{5}a_{6}}\mathring{D}\dot{K}^{a_{7}a_{8}}%
\end{align*}
\textbf{Theorem: \ The teleparallel equivalent of Euler Classes are closed
forms.}

Proof: \ First consider the case of the teleparallel equivalent of Euler class
in two dimensions, which is given by%

\begin{equation}
e_{2}(M)\sim-\left\langle \Gamma\right\rangle =-\varepsilon_{a_{1}a_{2}}%
\Gamma^{a_{1}a_{2}}=-\left\langle d\dot{K}+\dot{K}\dot{K}\right\rangle
,\nonumber
\end{equation}
where $\left\langle {}\right\rangle $ denotes the symmetrized trace and where
$\Gamma^{a_{1}a_{2}}=\mathring{D}K^{a_{1}a_{2}}+\left(  \dot{K}\dot{K}\right)
^{a_{1}a_{2}}$, or $\Gamma^{a_{1}a_{2}}=DK^{a_{1}a_{2}}-\left(  \dot{K}\dot
{K}\right)  ^{a_{1}a_{2}}$. \ So that,  $\Gamma=\varepsilon_{a_{1}a_{2}}%
\Gamma^{a_{1}a_{2}}=\mathring{D}\left(  \varepsilon_{a_{1}a_{2}}\dot{K}%
^{a_{1}a_{2}}\right)  +\varepsilon_{a_{1}a_{2}}\left(  KK\right)  ^{a_{1}%
a_{2}}=$ $\left\langle d\dot{K}+\dot{K}\dot{K}\right\rangle $. \ Therefore,%

\begin{align*}
d\left\langle \Gamma\right\rangle  &  =d\left\langle d\dot{K}+\dot{K}\dot
{K}\right\rangle =\left\langle d\left(  \dot{K}\dot{K}\right)  \right\rangle
\\
&  =\left\langle d\dot{K}\dot{K}-\dot{K}d\dot{K}\right\rangle =\left\langle
\left(  \Gamma-\dot{K}\dot{K}\right)  \dot{K}-\dot{K}\left(  \Gamma-\dot
{K}\dot{K}\right)  \right\rangle \\
&  =\left\langle \Gamma\dot{K}-\dot{K}\dot{K}\dot{K}-\dot{K}\Gamma+\dot{K}%
\dot{K}\dot{K}\right\rangle =\left\langle \Gamma\dot{K}-\dot{K}\Gamma
\right\rangle =0,
\end{align*}
where in the last step we have used a property of the symmetric trace.

Let us now consider the teleparallel equivalent for Euler's class in $2n$
dimensions, which is given by%

\begin{align}
e_{2n}(M) &  \sim\left(  -1\right)  ^{n}\left\langle \left(  \dot{K}\dot
{K}+\mathring{D}\dot{K}\right)  \left(  \dot{K}\dot{K}+\mathring{D}\dot
{K}\right)  \cdot\cdot\cdot\cdot\left(  \dot{K}\dot{K}+\mathring{D}\dot
{K}\right)  \right\rangle \nonumber\\
&  =\left(  -1\right)  ^{n}\left\langle \left(  \dot{K}\dot{K}+\mathring
{D}\dot{K}\right)  ^{n}\right\rangle =\left(  -1\right)  ^{n}\left\langle
\Gamma^{n}\right\rangle .\label{eu3'}%
\end{align}
This form is a closed form. In fact,
\begin{align*}
d\left\langle \Gamma^{n}\right\rangle  &  =\left\langle n\Gamma^{n-1}%
d\Gamma\right\rangle =\left\langle n\Gamma^{n-1}\left(  \Gamma\dot{K}-\dot
{K}\Gamma\right)  \right\rangle \\
&  =n\left\langle \Gamma^{n}\dot{K}-\Gamma^{n-1}\dot{K}\Gamma\right\rangle =0.
\end{align*}
So that%
\begin{equation}
de_{2n}(M)=0.\label{eu3''}%
\end{equation}

Let us now show that (\ref{eu3'}) is gauge invariant. In fact, the variation
of $e_{2n}(M)$ is given by%
\begin{align*}
\delta e_{2n}(M) &  \sim\left(  -1\right)  ^{n}\left\langle \delta\Gamma
^{n}\right\rangle =\left(  -1\right)  ^{n}\left\langle \left(  \dot{K}\dot
{K}+\mathring{D}\dot{K}\right)  ^{n}\right\rangle \\
&  =\left(  -1\right)  ^{n}n\left\langle \left(  \dot{K}\dot{K}+\mathring
{D}\dot{K}\right)  ^{n-1}\right\rangle \delta\left(  \dot{K}\dot{K}%
+\mathring{D}\dot{K}\right)  \\
&  =\left(  -1\right)  ^{n}n\left\langle \left(  \dot{K}\dot{K}+\mathring
{D}\dot{K}\right)  ^{n-1}\right\rangle \left(  2\dot{K}\delta\dot{K}%
+\mathring{D}\delta\dot{K}\right)  .
\end{align*}

Since, $\dot{T}^{\text{ }a}=\dot{K}_{\text{ \ }c}^{a}h^{c}$, $\dot{K}_{\text{
\ }c}^{a}=\dot{T}^{\text{ }a}h_{c}$, $\dot{T}^{\text{ }a}=\dot{D}h^{a}$ and
$\delta h^{a}=0$, we have $\delta\dot{T}^{\text{ }a}=0.$ \ So that $\delta
\dot{K}_{\text{ \ }c}^{a}=0$ and therefore%
\begin{equation}
\delta e_{2n}(M)\sim\left(  -1\right)  ^{n}\delta\Gamma^{n}=0.\label{eu'''}%
\end{equation}
This means that the teleparallel equivalent of the Euler topological invariant%
\[
\mathcal{P}_{2n}=\langle\Gamma^{n}\rangle,
\]
is a 2n-form satisfying the condition $d\mathcal{P}_{2n}=0$. According to the
Poincar\'{e} lemma, we can write $\mathcal{P}_{2n}=d\mathcal{C}_{2n-1}$, where
$\mathcal{C}_{2n-1}$ is a $\left(  2n-1\right)  $-form that we will call the
teleparallel equivalent of the Chern- Simons form.

\section{\textbf{Teleparallel Equivalent of Lanczos-Lovelock Action}}

We have seen that the Lanczos-Lovelock action is given by a linear combination
of all the dimensionally continued Euler classes of dimension $2p<d$. Let us
now consider the construction of the teleparallel equivalent of the
$2p$-dimensional Euler classes dimensionally continued to $d>2p$ dimensions.

\begin{enumerate}
\item[(a)] The two-dimensional Euler class dimensionally continued to $d$
dimensions is given by%
\begin{align*}
e_{2dc}(M)  &  \equiv L^{(1)}\sim\varepsilon_{a_{1}a_{2}a_{3\cdot\cdot
\cdot\cdot}a_{d}}\dot{K}_{\text{ \ \ }c_{1}}^{a_{1}}\dot{K}^{c_{1}a_{2}%
}h^{a_{3}}\cdot\cdot\cdot\cdot h^{a_{d}}\\
&  =\varepsilon_{a_{1}a_{2}a_{3\cdot\cdot\cdot\cdot}a_{d}}\left(  \dot{K}%
^{2}\right)  ^{a_{1}a_{2}}h^{a_{3}}\cdot\cdot\cdot\cdot h^{a_{d}}%
\end{align*}

\item[(b)] For the four-dimensional Euler class dimensionally continued to $d$
dimensions we have
\end{enumerate}

\[
e_{4dc}(M)\equiv L^{(2)}\sim\varepsilon_{a_{1}a_{2}a_{3\cdot\cdot\cdot\cdot
}a_{d}}\left(  2\dot{K}^{a_{1}a_{2}}D\dot{K}_{\text{ \ }c_{3}}^{a_{3}}\dot
{K}^{c_{3}a_{4}}+\left(  \dot{K}^{2}\right)  ^{a_{1}a_{2}}\left(  \dot{K}%
^{2}\right)  ^{a_{3}a_{4}}\right)  h^{a_{5}}\cdot\cdot\cdot\cdot h^{d}%
\]

\begin{enumerate}
\item[(c)] The $2p$-dimensional Euler class dimensionally continued to $d$
dimensions is similarly given by
\end{enumerate}

\begin{align*}
e_{2pdc}(M)  &  \equiv L^{(p)}\sim\left(  -1\right)  ^{p}\varepsilon
_{a_{1}a_{2}\cdot\cdot\cdot\cdot a_{d}}\left(  \dot{K}\dot{K}+\mathring{D}%
\dot{K}\right)  ^{a_{1}a_{2}}\left(  \dot{K}\dot{K}+\mathring{D}\dot
{K}\right)  ^{a_{3}a_{4}}\cdot\cdot\cdot\\
&  \cdot\cdot\cdot\left(  \dot{K}\dot{K}+\mathring{D}\dot{K}\right)
^{a_{2p-1}a_{2p}}h^{a_{2p+1}}\cdot\cdot\cdot\cdot h^{d}%
\end{align*}

\textbf{Teleparallel equivalent of Lovelock gravity action:} \ we define the
teleparallel equivalent of the Lanczos-Lovelock gravity action by a linear
combination of dimensionally continued Euler classes. This action can be
written in the form%

\begin{equation}
S=\int\sum_{p=0}^{\left[  \frac{D}{2}\right]  }\alpha_{p}L^{\left(  p\right)
}, \label{L1}%
\end{equation}
where
\begin{equation}
L^{(p)}\sim\left(  -1\right)  ^{p}\varepsilon_{a_{1}a_{2}\cdot\cdot\cdot\cdot
a_{d}}\left(  \dot{K}\dot{K}+\mathring{D}\dot{K}\right)  ^{a_{1}a_{2}}\left(
\dot{K}\dot{K}+\mathring{D}\dot{K}\right)  ^{a_{3}a_{4}}\cdot\cdot\cdot
\cdot\left(  \dot{K}\dot{K}+\mathring{D}\dot{K}\right)  ^{a_{2p-1}a_{2p}%
}h^{a_{2p+1}}\cdot\cdot\cdot\cdot h^{d}, \label{81'}%
\end{equation}
and [($d/2$)] is the integer part of ($d/2$). The coefficients $\alpha_{p}$,
with $p=0,1,\cdot\cdot\cdot,\left[  d/2\right]  $, are arbitrary real
constants that are not fixed from first principles.

From (\ref{L1}, \ref{81'}) we can see that: $(i)$ for $p=1$ we obtain the
equations $(30)$ and $(23)$ of references \cite{kof,gon}; $\left(  ii\right)
$ for $p=2$ we obtain the results given by the equations $(47)$ and $(35)$ of
references \cite{kof} and \cite{gon}; and $\left(  iii\right)  $ for $p=p$ we
obtain the result given by the equation $(54)$ of Ref. \cite{gon} when we
replace $\mathring{D}\dot{K}$ by $\dot{D}\dot{K}-2\dot{K}\dot{K}.$

On the other hand, following the same produce it is possible to obtain the
teleparallel equivalent of the actions (\ref{a1}, \ref{a2}, \ref{a3} )
introduced in Ref. \cite{criso}. In fact, for the action (\ref{a1}) we find
\[
S_{2}=\kappa\int\varepsilon_{a_{1}a_{2}\cdot\cdot\cdot\cdot a_{d}}%
(\frac{l^{-4}}{d}e^{a_{1}}\cdot\cdot\cdot e^{a_{d}}+\frac{2l^{-2}}{d-2}\left(
\dot{K}\dot{K}+\mathring{D}\dot{K}\right)  ^{a_{1}a_{2}}h^{a_{3}}\cdot
\cdot\cdot h^{a_{d}}%
\]%
\begin{equation}
+\frac{1}{d-4}\left(  \dot{K}\dot{K}+\mathring{D}\dot{K}\right)  ^{a_{1}a_{2}%
}\left(  \dot{K}\dot{K}+\mathring{D}\dot{K}\right)  ^{a_{3}a_{4}}h^{a_{5}%
}\cdot\cdot\cdot h^{a_{d}}),\label{a1'}%
\end{equation}
which exists only for $d>4.$

For the action (\ref{a2}) we have
\[
S=\int\sum_{p=0}^{\left[  d/2\right]  }\frac{\kappa}{d-2p}\binom{n-1}%
{p}l^{2p-d+1}\varepsilon_{a_{1}a_{2}\cdot\cdot\cdot\cdot\cdot\cdot a_{d}%
}\left(  \dot{K}\dot{K}+\mathring{D}\dot{K}\right)  ^{a_{1}a_{2}}\cdot
\cdot\cdot
\]%
\begin{equation}
\cdot\cdot\cdot\cdot\left(  \dot{K}\dot{K}+\mathring{D}\dot{K}\right)
^{a_{2p-1}a_{2p}}h^{a_{2p+1}}\cdot\cdot\cdot\cdot h^{a_{d}} \label{a2'}%
\end{equation}

Finally, for the action (\ref{a3}) the parallel equivalent is given by
\[
S=\int\sum_{p=0}^{\left[  d/2\right]  }\frac{\kappa}{2n}\binom{n}{p}%
l^{2p-d+1}\varepsilon_{a_{1}a_{2}\cdot\cdot\cdot\cdot\cdot\cdot a_{d}}%
^{a_{1}a_{2}}\left(  \dot{K}\dot{K}+\mathring{D}\dot{K}\right)  ^{a_{1}a_{2}%
}\cdot\cdot\cdot
\]%
\begin{equation}
\cdot\cdot\cdot\left(  \dot{K}\dot{K}+\mathring{D}\dot{K}\right)
^{a_{2p-1}a_{2p}}h^{a_{2p+1}}\cdot\cdot\cdot\cdot h^{a_{d}}. \label{a3'}%
\end{equation}

From (\ref{eu'''}) we can see that the actions (\ref{L1}, \ref{a1'},
\ref{a2'}, \ref{a3'}) are invariant under gauge translations. This means that
these actions are invariants under Poincar\'{e} group and diffeomorphism.

\section{Concluding Remarks}

In this article we have found that the teleparallel equivalent of the
$d$-dimensional Euler classes and then it is shown that the teleparallel
equivalent of the Lanczos-Lovelock gravity action is generated by dimensional
continuation of the Euler characteristics associated to all the lower even
dimensions, which are invariant under the Poincare group as well as under diffeomorphisms.

\begin{acknowledgement}
This work was supported in part by FONDECYT grants 1130653 and 1150719 from
the Government of Chile. One of the authors (NA) was suported by grants
22140902 from CONICYT (National Commission for Scientific and Technological
Re-search, spanish initials) and from Universidad de Concepci\'{o}n, Chile.
\end{acknowledgement}

\bigskip

-------------------------------------------------------------------

\section{Appendix A: Linear Frames and Tetrads}

We are going to use the greek alphabet ($\mu,\nu,\rho,\cdot\cdot\cdot
=0,1,2,3$) to denote indices related to spacetime, and the first half of the
latin alphabet $\left(  a,b,c,\cdot\cdot\cdot=0,1,2,3\right)  $ to denote
indices related to the tangent space, a Minkowski spacetime whose Lorentz
metric is assumed to have the form $\eta_{ab}=diag\left(  +1,-1,-1,-1\right)
.$

The middle letters of the latin alphabet $\left(  i,j,k,\cdot\cdot
\cdot=0,1,2,3\right)  $ will be reserved for space indices. A general
spacetime is a $4$-dimensional differential manifold $%
\mathbb{R}
^{3,1},$ whose tangent space is, at any point, a Minkowski spacetime.

Spacetime coordinates will be denoted by $\left\{  x^{\mu}\right\}  $, whereas
tangent space coordinates will be denoted by $\left\{  x^{a}\right\}  $. Such
coordinate systems determine, on their domains of definition, local bases for
vector fields, formed by the sets of gradients $\left\{  \partial_{\mu
}\right\}  $ \ \ and \ \ $\left\{  \partial_{a}\right\}  ,$ as well as the
dual bases $\left\{  dx^{\mu}\right\}  $ and $\left\{  dx^{a}\right\}  $ for
$1$-forms fields.

A holonomic base (or coordinate base) like $\left\{  e_{a}\right\}  =\left\{
\partial_{a}\right\}  $, related to a coordinate system, is a very particular
case of a linear base. Notice that, on a general manifold, vector fields are
(like coordinate systems) only locally defined.

A frame field provides, at each point $p\in%
\mathbb{R}
^{3,1}$, a base for the vectors on the tangent space $T_{p}%
\mathbb{R}
^{3,1}$. Of course, on the common domains they are defined, the members of a
base can be written in terms of the members of the other. For example,
$e_{a}=e_{a}^{\text{ \ }\mu}\partial_{\mu}$ and $e^{a}=e_{\text{ \ }\mu}%
^{a}dx^{\mu}$ and conversely $\partial_{\mu}=e_{\text{ \ }\mu}^{a}e_{a}$ \ and
\ $dx^{\mu}=e_{a}^{\text{ \ }\mu}e^{a}$.

Consider now the spacetime metric $\mathbf{g}$, with components $g_{\mu\nu}$ ,
in some dual holonomic base $\left\{  dx^{\mu}\right\}  $: $\mathbf{g=}%
g_{\mu\nu}dx^{\mu}\otimes dx^{\nu}.$ A tetrad field $h_{a}=h_{a}^{\text{
\ }\mu}\partial_{\mu}$ \ will be a linear base which relates $\mathbf{g}$ to
the tangent space metric $\eta=\eta_{ab}dx^{a}\otimes dx^{b}=\eta_{ab}%
dx^{a}dx^{b}$ by $\eta_{ab}=g\left(  h_{a},h_{b}\right)  =g_{\mu\nu}%
h_{a}^{\text{ \ }\mu}h_{b}^{\text{ \ }\nu}.$

The components of the dual base members $h^{a}=h_{\text{ }\nu}^{a}dx^{\nu}$
satisfy
\begin{equation}
h_{\text{ }\mu}^{a}h_{a}^{\text{ \ }\nu}=\delta_{\mu}^{\nu}\text{ \ \ and
\ }h_{\text{ }\mu}^{a}h_{b}^{\text{ \ }\mu}=\delta_{b}^{a}\text{\ }\label{Ap1}%
\end{equation}
so that $g_{\mu\nu}=\eta_{ab}h_{\text{ }\mu}^{a}h_{\text{ }\nu}^{b}.$ Since%

\begin{equation}
g=\det\left(  g_{\mu\nu}\right)  =\det\left(  \eta_{ab}h_{\text{ }\mu}%
^{a}h_{\text{ }\nu}^{b}\right)  ,\label{Ap2}%
\end{equation}
we can write%
\begin{equation}
h=\det\left(  h_{\text{ }\mu}^{a}\right)  =\sqrt{-g}.\label{Ap3}%
\end{equation}

Anholonomy is the property by which a differential form is not the
differential of anything, or of a vector field which is not a gradient of
anything. Consider a dual base $h^{a}$ such that $dh^{a}\neq0$, that is, not
formed by differentials. Applying the anholonomic 1-forms $h^{a}$ to
$\partial_{\mu}$, we obtain, $h_{\text{ \ }\mu}^{a}=h^{a}\left(  \partial
_{\mu}\right)  $, which lead to the components of each  $h^{a}=h_{\text{
\ }\mu}^{a}dx^{\mu}$ along $dx^{\mu}$. \ 

An anholonomic basis (or non-coordinate basis) $\left\{  h_{a}\right\}  $
satisfies the commutation relation%

\begin{equation}
\left[  h_{a},h_{b}\right]  =f_{\text{ }ab}^{c}h_{c}\label{Ap4}%
\end{equation}
where $f_{\text{ }ab}^{c}$ are the so-called structure constans or
coefficients of anholonomy.

The dual expression of the commutation relation above is the so called Cartan
structure equation%

\begin{equation}
dh^{c}=-\frac{1}{2}f_{\text{ }ab}^{c}h^{a}\wedge h^{b}=\frac{1}{2}\left(
\partial_{\mu}h_{\text{ \ }\nu}^{c}-\partial_{\nu}h_{\text{ \ }\mu}%
^{c}\right)  dx^{\mu}dx^{\nu}.\label{Ap5}%
\end{equation}

The structure coeffcients represent the curls of the base members:%

\begin{equation}
f_{\text{ }ab}^{\text{ }c}=h_{a}^{\text{ }\mu}h_{b}^{\text{ }\nu}\left(
\partial_{\mu}h_{\text{ \ }\nu}^{c}-\partial_{\nu}h_{\text{ \ }\mu}%
^{c}\right)  .\label{Ap6}%
\end{equation}

Notice that when $f_{\text{ }ab}^{c}=0$ we have $dh^{a}=0$. This means that
$h^{a}$ is a closed differential form.

In the presence of gravitation, $f_{\text{ }ab}^{c}$ includes both inertial
and gravitational effects. In this case, the spacetime metric represents a
general (pseudo) riemannian spacetime. On the other hand, in absence of
gravitation,  the anholonomy of the frames is entirely related to the inertial
forces present in those frames. In this case, $h_{\text{ \ }\mu}^{a}$ becomes
trivial and $g_{\mu\nu}$ turns out to represent the Minkowski metric in a
general coordinate system. In this case, the $h_{\text{ \ }\mu}^{a}$ are
called holonomic frames.

\section{Appendix B}

\subsection{\textbf{General Covariance Principle}}

The general covariance principle\textbf{ }\cite{cov1}\textbf{ }states that a
valid equation in special relativity can still be valid in the presence of
gravitation if it is generally covariant, i.e., if it preserves its shape
under general coordinate transformations. To make an equation generally
covariant, it is necessary to introduce a connection which, in principle, is
related only to the inertial properties of the coordinate system.

In the case that only transformations of coordinates are involved, then a
vacuum connection is enough, that is to say with null curvature and zero
torsion. The use of the equivalence between inertial and gravitational effects
allows replacing the (vacuum) connection with a connection that represents a
true gravitational field.

The important thing here is to keep in mind that: the equations have the same
form for both empty and non-empty connections. This means that the valid
equations in the presence of gravitation are obtained from the corresponding
valid equations in special relativity.

This means that the principle of general covariance can be understood as an
active version of the principle of equivalence in the sense that, by making a
special-relativistic equation covariant and then using the strong equivalence
principle, it is possible to obtain the form of such equation in the presence
of gravitation.

On the other hand, the usual form of the equivalence principle can be
interpreted as the passive version of this principle: the special relativistic
equation must be recovered in a locally inertial reference system.

It should be emphasized that general covariance by itself is empty of physical
content, in the sense that any equation can be made generally covariant. Only
when the local equivalence between inertial and gravitational effects is used,
and the compensating term is re-interpreted as representing a true
gravitational field, the general covariance principle can be seen as an active
version of the principle of strong equivalence \cite{cov2}\textbf{.}

The description we have made up to now of the covariance principle is known as
the holonomic version of this fundamental principle. An alternative and more
general version of the general covariance principle can be obtained using
non-holonomic frames. The fundamental difference between the two versions is
that, instead of requiring an equation to be covariant under a general
coordinate transformation (holonomic frames), in the anholonomic-frame version
the equation is required to be covariant under a local Lorentz transformation
of the frame. The physical content of this version is the same as that of the
holonomic version \cite{cov3}\textbf{.}

The non-holonomic (non-coordinate) version of the general covariance principle
is more general than its holonomic (coordinate) version, because the
non-holonomic version is valid for both integer spin fields and half-integer
spin fields.

The importance of the general covariance principle lies in the fact that this
principle defines in a natural way a Lorentz covariant derivative, and
therefore also defines a gravitational coupling procedure. The procedure for
obtaining the coupling consists of two steps.

\begin{itemize}
\item The first step is to pass to a general anholonomic frame, where inertial
effects appear in the form of a compensating term, where the inertial effects,
which appear in the form of an empty Lorentz connection, are present.

\item The second step is to use the  strong equivalence principle to replace
the inertial effects given by an empty connection by a connection that
represents a true gravitational field. This gives rise to the process of
gravitational coupling.
\end{itemize}

\subsection{Coupling Procedure}

In gauge theories, the coupling prescription amounts to replace ordinary
derivatives by covariant derivatives involving a connection. As an example,
let us consider a gauge theory for the unitary group $U(1)$. Under an
infinitesimal gauge transformation with parameter $\alpha=\alpha(x)$, a spinor
field changes according to $\delta\psi=\psi^{\prime}-\psi\approx i\alpha\psi$.
The ordinary derivative $\partial_{\mu}\psi$ transform under $U(1)$ as
$\delta\left(  \partial_{\mu}\psi\right)  =i\alpha\partial_{\mu}\psi+i\left(
\partial_{\mu}\alpha\right)  \psi.$ This means does not transform covariantly,
i.e., does not transform as a field. \ 

In order to recover the covariance, that is to say that the field derivative
transforms as a field, it is necessary to introduce a gauge potential $A_{\mu
}$, which is a connection that takes values in the Lie algebra of the gauge
group $U(1)$.

The covariant derivative given by
\begin{equation}
D_{\mu}\psi=\partial_{\mu}\psi+iA_{\mu}\psi,\label{14}%
\end{equation}
transforms as
\begin{equation}
\delta\left(  D_{\mu}\psi\right)  =i\alpha\left(  D_{\mu}\psi\right)
,\label{15}%
\end{equation}
i.e., transform as a field provided that the gauge potential transforms as an
abelian connection, i.e., as%
\begin{equation}
\delta A_{\mu}=-\partial_{\mu}\alpha.\label{16}%
\end{equation}
This tells us that the coupling procedure consists of replacing the ordinary
derivative with a covariant derivative%
\begin{equation}
\partial_{\mu}\psi\longrightarrow D_{\mu}\psi.\label{17}%
\end{equation}

On the other hand, due to the fact that gravitation is not a background
independent theory, the gravitational coupling procedure has two distinct parts.

\begin{itemize}
\item The first part is to replace the Minkowski metric $\eta_{\mu\nu}$ by a
general pseudoriemannian metric $g_{\mu\nu}$ representing a gravitational
field: \ $\eta_{\mu\nu}\longrightarrow g_{\mu\nu}.$ This part is universal in
the sense that it affects equally all matter fields. This follows naturally
from the requirement of covariance under space-time translations.

\item The second part is related to the coupling of the spin of the matter
fields to the gravitational field, and is related to the requirement of
covariance under Lorentz transformations. The coupling appears in the form
$\partial_{\mu}\longrightarrow D_{\mu},$ where $D_{\mu}$ is a Lorentz
covariant derivative. Obviously this part of the coupling is not universal in
the sense that it depends on the spin content of each field. 

Let us consider these two coupling procedures separately. \ 
\end{itemize}

\textbf{Translational Coupling: }Under an infinitesimal gauge translation, the
field $\psi$ transforms according
\[
\delta\psi=\psi^{\prime}-\psi=\varepsilon\psi=\varepsilon^{a}\partial_{a}\psi,
\]
and its ordinary derivative $\partial_{\mu}\psi$ transform as%

\begin{equation}
\delta\left(  \partial_{\mu}\psi\right)  =\varepsilon^{a}\partial_{a}\left(
\partial_{\mu}\psi\right)  +\left(  \partial_{\mu}\varepsilon^{a}\right)
\partial_{a}\psi, \label{21}%
\end{equation}
that is, the derivative of the field does not transform as a field.

In order to recover covariance, it is necessary to introduce a gauge potential
$e_{\mu}$, a $1$-form assuming values in the Lie algebra of the translation
group: $e_{\mu}=e_{\mu}^{\text{ }a}P_{a}$, where $P_{a}=\partial_{a}$ are the
infinitesimal translation generators which satisfy $\left[  P_{a}%
,P_{b}\right]  =0.$ \ 

A gauge transformation is defined as a local translation of the tangent-space
coordinates, $x^{\prime a}=x^{a}+\varepsilon^{a}$, where $\varepsilon
^{a}=\varepsilon^{a}(x^{\mu})$ are the corresponding infinitesimal parameters.
In terms of $P_{a},$ it can be written in the form $\delta x{}^{a}%
=\varepsilon^{b}P_{b}\,x^{a}.$ \ Using the general definition of covariant
derivative, we have%

\begin{equation}
D\psi=(d+e)\psi=(d+e^{a}P_{a})\psi.\label{tel2}%
\end{equation}
From (\ref{tel2}) we can see
\begin{equation}
D_{\mu}=\left(  \partial_{\mu}x^{a}+e_{\mu}^{\text{ }a}\right)  \partial
_{a}\psi=D_{\mu}x^{a}\partial_{a}\psi=h_{\mu}^{\text{ }a}\partial_{a}%
\psi,\label{tel3}%
\end{equation}
where%

\begin{equation}
h_{\mu}^{\text{ \ }a}=\partial_{\mu}x^{a}+e_{\mu}^{\text{ \ }a}\text{ or
}h^{\text{ }a}=Dx^{a}=dx^{a}+e^{a}\label{tel5}%
\end{equation}
is a nontrivial, (anholonomic)  tetrad field. In fact, under a translation,
the gauge potential transforms as an abelian connection, i.e., as
\begin{equation}
\delta e^{a}=d\varepsilon^{a}\label{tel6}%
\end{equation}
and, given that $\delta x^{a}=-\varepsilon^{a},$ we have
\begin{equation}
\delta h^{a}=0.\label{tel7}%
\end{equation}

If we write the ordinary derivative $\partial_{\mu}\psi$ as%

\begin{equation}
\partial_{\mu}\psi=e_{\text{ }\mu}^{\text{ \ }a}\partial_{a}\psi\label{33}%
\end{equation}
with $e_{\text{ }\mu}^{a}=\partial_{\mu}x^{a}$, the translational coupling
prescription takes the form
\begin{equation}
e_{\text{ }\mu}^{a}\partial_{a}\psi\longrightarrow h_{\mu}^{a}\partial_{a}%
\psi.\label{34}%
\end{equation}
This means that
\begin{equation}
e_{\text{ }\mu}^{a}\longrightarrow h_{\mu}^{a}.\label{35}%
\end{equation}

Concomitantly with this replacement, the space-time metric changes according
to
\begin{equation}
\eta_{\mu\nu}=\eta_{ab}e_{\text{ }\mu}^{a}e_{\text{ }\mu}^{b}\longrightarrow
g_{\mu\nu}=\eta_{ab}h_{\text{ }\mu}^{a}h_{\text{ }\mu}^{b}.\label{36}%
\end{equation}

The change in the spacetime metric is therefore, a direct consequence of the
translational coupling prescription.
\end{document}